# Solar Thermochemical Water-Splitting Reaction Enhanced by Hydrogen Permeation Membrane

*Chenxi Sui[2], Hongsheng Wang[1*], Xiang Liu[1] and Xuejiao Hu[1*]*

[1] School of Power and Mechanical Engineering, Wuhan University, Wuhan, Hubei 430072, P. R. China

[2] School of Physics and Technology, Wuhan University, Wuhan, Hubei 430072, P. R. China

*Email address: wanghongsheng@whu.edu.cn, xjhu@whu.edu.cn

**Abstract**

The low conversion rate and efficiency always weaken the performance of thermochemical water-splitting reaction. Herein, we, for the first time, conducted the thermodynamic study of a hydrogen permeable membrane (HPM) in an isothermal thermochemical water-splitting reaction driven by solar energy, which has showed a sharply enhanced conversion rate of 87.8% at 1500 $^o$C and $10^{-5}$ atm at permeated side (versus 1.26% with oxygen permeation membrane). According to thermodynamic analysis, the first-law thermodynamic efficiency can reach as high as 59.1%. When taking solar-to-electric efficiency and vacuum pump efficiency into account (for converting separation work into solar energy), we simulated the appreciable efficiency of 3.05% at 1500 $^o$C. The numerical model will provide guidance for the actual production of hydrogen by high temperature solar water splitting. Such novel work manifests the great significance of constructing a HPM reactor for efficient solar thermochemical water splitting, which shows a novel approach for high-temperature solar water splitting.

# 1. Introduction

With the shortage of clean and sustainable energy, solar energy plays an important role in meeting the global energy needs, for its extensive distribution in the world and inexhaustible nature. Solar energy is difficult to store, transport, and discontinuous in space and time, so it needs to be converted into fuel that can be used continuously and, stored and transported conveniently. It remains challenging to covert the solar radiation into portable and preservable energy efficiently. Despite the already appeared photovoltaic chemistry water splitting, the effective use of the entire solar spectrum has not yet been realized in such system [1]. Therefore, thermochemistry water splitting by utilizing the all-spectrum solar radiation has gone to the spotlight [2-6]. Among these exciting researches, one of the main methods of thermochemical water splitting is two-step thermochemical cycles [6, 7] driven by solar energy, because they bypass the $H_2$-$O_2$ separation problem [6]. However, such method has a huge temperature difference (~800 °C) and always suffers from the difficulties of solid-state sensible heat (including catalyst, reaction chamber insulation, etc.) recovery [13], which leads to the experimental efficiency of the dual-temperature method only a few thousandths [6], and materials damage induced by thermal stress [14]. In order to avoid these disadvantages, a two-step isothermal water splitting cycle process has been proposed [5, 9-11], which changes the oxygen partial pressure between oxidation step and reduction step instead of temperature swing, but the discontinuous solar heat utilization (only reduction step requires high temperature heat) decrease the efficiency of solar energy utilization. Hao group demonstrated that the solar energy can be used continuously by constructing isothermal oxygen permeation membrane (OPM) reactor [15]. Nevertheless, the most important factor in solar fuel production is the solar-to-fuel efficiency. Both dual-temperature and OPM isothermal methods are very inefficient [5, 9]. This is be proved by thermodynamic analysis. According to the theory of thermochemistry, the chemical equilibrium constant of water splitting reaction can be calculated as:

$$K(T) = \frac{(P_{H_2}/P^\ominus) \cdot (P_{O_2}/P^\ominus)^{0.5}}{P_{H_2O}/P^\ominus} \tag{1}$$

where $P_{H_2}$, $P_{O_2}$ and $P_{H_2O}$ are the partial pressures of $H_2$, $O_2$ and $H_2O$, respectively, $P^\ominus$ is standard pressure. It is obvious that the conversion rate of water splitting is lower for separation of $O_2$ than $H_2$ under the same partial pressure outside of the membrane, due to Le Chatelier's Principle.

It has been demonstrated in the published works that, the theoretical conversion ratio limit of the isothermal method (including two-step isothermal cycle and OPM reactor) is only 1.26% at partial pressure outside the membrane of $10^{-5}$ atmospheres (atm) for oxygen separation [6, 15]. Furthermore, the thermodynamic efficiency of solar-to-fuel conversion can be expressed as

$$\eta = \frac{n_{H_2} \cdot HHV_{H_2}}{Q_{solar}} \qquad (2)$$

where $HHV_{H_2}$ is the higher heating value of $H_2$, $n_{H_2}$ is the amount of produced $H_2$ and $Q_{solar}$ is the solar energy input to produce $H_2$ due to the disassociation of $H_2O$. Axiomatically, the lower the conversion rate, the more reactants need to be consumed to produce the same amount of hydrogen, which will increase the energy consumption and thus lead to a lower efficiency.

Hence, we must start from the nature of chemical reaction and explore ways to increase the efficiency of isothermal methods. According to the above-mentioned thermodynamic analysis, separation of as-formed gas ($H_2$ or $O_2$) can improve the conversion rate and therefore elevate efficiency. Follow this road, many researchers use oxygen permeation membrane (OPM) to separate oxygen because of the intriguing kinetic and mechanical properties [5, 9, 15, 26]. However, because the stoichiometric amount of oxygen is only half that of hydrogen, the partial pressure of hydrogen is twice of that of oxygen on the reaction side. By giving the equal negative pressure outside membrane to oxygen and hydrogen, more amount of hydrogen is separated by HPM, which means a higher conversion rate of water by HPM reactor. Thus, efficiency has not been fully improved with the utilization of OPM. Therefore, if the $H_2$ could be separated by using hydrogen permeation membrane (HPM), compared with the separation of oxygen under the same partial pressure outside the membrane, the chemical equilibrium will shift forward further and the conversion rate will increase further, and the efficiency will be further improved. Herein, we, for the first time, put up a thermodynamic analysis of an isothermal water splitting system by using HPM and the conversion rate and efficiency were improved dramatically. According to thermodynamic relationship, the reaction temperature must be higher than 955 °C otherwise the conversion rate of water vapor would be lower than $9.67 \times 10^{-6}$%, and the partial pressures of hydrogen would be lower than $10^{-5}$ atm, which is too low for practical application with physical separation method (vacuum pumping, carrier gas purging, etc.). In order to make the system practical (the reaction condition must simultaneously meet the temperature tolerance of the HPM and high system efficiency), there are two ways to optimize the HPM system: i) find the HPM which can endure the temperature higher than 955 $^{o}$C with a good permeation performance; ii) reduce the reaction temperature under the premise of ensuring the efficiency of water splitting.

As for the materials of HPM, there are many types and numbers available [16]. However, the main challenge is that most of the HPM is rarely used at the temperature higher than 1000$^{o}$C [17-20]. According to the thermodynamic analysis above, the simultaneous separation of hydrogen with HPM and increase of positive pressure $P_0$ (reactants feeding pressure) can elevate the amount of permeated hydrogen and therefore shift the equilibrium forward and enhance the conversion rate. This is because that the pressure difference between the inside and outside of the membrane increases, the amount of hydrogen separated increases, and the chemical equilibrium moves in the positive direction. Encouraged by this, the reaction temperature can be reduced with a relatively lighter compromise of solar-to-fuel efficiency. Simultaneously, it is worthwhile mentioning that some HPM materials, such as metal palladium and perovskite, could stand the temperature over 900$^{o}$C [19, 24], and $CeO_2$ and doped $CeO_2$ ( $(CeO_2)_{0.9}(GdO_{1.5})_{0.1}$ and $Ce_{0.8}Yb_{0.2}O_{1.9}$ ) is even able to stand the temperature as high as 1527$^{o}$C [21-23]. Thus, they are able to be used as HPM for high-temperature water splitting and, hereby, Pd membrane was introduced in reactor for its rapid hydrogen penetration rate [19] and high melting point about 1555$^{o}$C [25]. Despite the fact that there are few studies on hydrogen permeability over 1000$^{o}$C in the published papers about metal palladium hydrogen permeation membranes, it might be because there is no such high temperature requirement in general. According to the higher melting point of the palladium material, we reasonably expand the useful temperature range of the palladium membrane in order to carry out the thermodynamic analysis of the HPM water splitting by using the palladium membrane as an example. It should be mentioned that there are many materials (Pd, doped $CeO_2$, zeolitic imidazolate framework and perovskite) [16, 19, 21-24] are available for hydrogen permeation at high temperature. These materials only have different dynamic properties and do not affect the thermodynamic efficiency limits of the entire system. We do not discuss the problem of the complex engineering technology of membrane materials. Thus, the main point is the system innovation of water splitting by hydrogen permeation membrane.

Furthermore, in traditional two-temperature thermodynamic cycle, oxygen atoms are 'dissolved' in ceria oxide solids [12], so it is impossible to exert a positive pressure inside of the solids to release oxygen atoms outward [6]. However, the HPM reactor could accomplish it. Adding positive pressure in HPM reactor can increase the pressure difference between the reactant side and permeation side of the membrane, increasing the amount of hydrogen permeability. Besides, increasing the pressure of water vapor can enhance the $H_2$ permeation rate, also make the equilibrium move forward and thus improve the conversion rate. In addition, because of the positive pressure, extremely low negative pressure is no longer needed and the too low negative pressure will lead to a sharp drop in vacuum pump efficiency, which can be expressed as $\eta_{pump} = \left(\frac{P_{H_2}}{P^{\ominus}}\right)^{0.544}$ [27, 28]. Besides, the reactant feeding pump efficiency is about 85% [31], which is much higher than vacuum pump efficiency at low pressure (e.g. ~0.2% for $10^{-5}$ atm). In the efficiency equation mentioned before (Eq. 2), $Q_{solar}$ on the denominator encompass the reactant feeding pump loss and the vacuum pump work loss which have been converted into the consumed solar radiation energy by using solar-to-electric conversion efficiency. Therefore, even if relying on a vacuum pump, it is possible to increase the negative pressure to improve the operating efficiency of the vacuum pump, reduce the solar energy loss ($Q_{solar}$), and improve the system energy utilization efficiency.

In this work, we constructed a theoretical model for characterization of the thermodynamic efficiency limit of water splitting with a HPM reactor operated under isothermal conditions with different positive (reactant feeding side) and negative (hydrogen permeated side) pressures. We will systematically discuss the influences of temperature and pressure and other key factors on the performance (gas partial pressure, conversion rate and thermodynamic efficiency) of the system in the following. This study will provide a novel idea and research platform for improving the efficiency of thermochemical decomposition of water.

## 2. Theoretical Calculation

The schematic demonstration of solar thermochemical HPM reactor is shown in Fig. 1, where the HPM tube is exposed to sunlight that is focused by the reflectors and can be heated to a high temperature for both activated permeation of $H_2$ and noticeable water splitting. The water is continuously fed into the HPM tube for thermolysis, while the chamber outside the HPM tube is kept at a low partial pressure of $H_2$ by vacuum pump (thereby the chemical potential difference between the HPM is high).

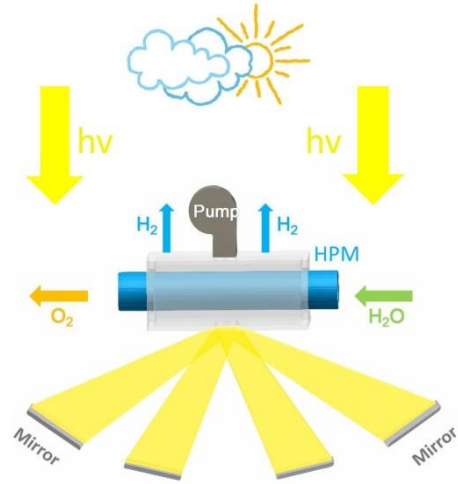

**Figure 1.** Schematic demonstration of a solar-driven hydrogen permeation membrane (HPM) reactor.

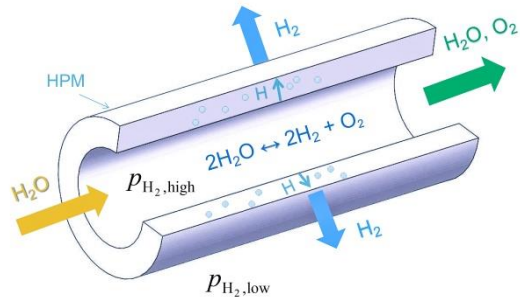

**Figure 2.** Cross section of HPM and the mechanism of its H$_2$ permeation.

The transportation of hydrogen across the palladium is partially because of the "jump" of hydrogen atoms through the octahedral interstitial sites of the face-centered cubic palladium lattice [29] and surface contact process [32]. For the demonstration of HPM tube of its H$_2$ permeation mechanism, the illustration of HPM is shown in Fig. 2. The ability of the transportation of H$_2$ through HPM is named as permeation flux. The flux of H$_2$ through the Pd membrane is expressed as [19]

$$J_{H_2} = -k \frac{(P_{H_2,ret}^{0.5} - P_{H_2,perm}^{0.5})}{X_M} \qquad (3)$$

where $P_{H_2,ret}$ is H$_2$ partial pressure inside (reaction side) the HPM tube, $P_{H_2,perm}$ is the partial pressure outside ($H_2$ permeation side) the HPM, $k$ is Sieverts constant and $X_M$ is the thickness of HPM. Specifically, $k$ can be fitted with an Arrhenius-type relation [19]

$$k \left[\text{mol}/(\text{m}\cdot\text{s}\cdot\text{Pa}^{0.5})\right] = 1.92\times10^{-7} \exp\left[-\frac{13810 \, (\text{J/mol})}{8.314 \, (\text{J/mol}\cdot\text{K})\times T \, (\text{K})}\right] \quad (4)$$

For the simplicity of numerical calculation, we defined the control volume along the HPM tube as a cylinder of length $dL$ according to the differential principle. $H_2$ inside the control volume is pulled away and can be characterized by $J_{H_2}$. Oxygen and steam are left inside, making the equilibrium of reaction shift forward to generate more hydrogen. The change rate of $P_{H_2,\text{ret}}$ (or $P_{O_2}$) inside HPM is decided by the function of themselves. That is, between any given time $t$ and next moment $t+\Delta t$, the amount of $O_2$ inside the membrane is $n_{O_2}(t)$ and $n_{O_2}(t+\Delta t)$, respectively. According to conservation law of mass, the amount of $O_2$ added in the same time is equal to half of the amount of pumped $H_2$. Such a relation can be expressed as

$$n_{O_2}(t+\Delta t) - n_{O_2}(t) = 0.5 J_{H_2} \cdot A_s \cdot \Delta t \quad (5)$$

where $A_s$ is the side area of the control volume (calculated by the logarithmic mean radius). The logarithmic mean radius is calculated as $R_m = (R_o - R_{in})/\ln(R_o/R_{in})$ [15]. Hence,

$$\frac{dn_{O_2}}{dt} = \lim_{\Delta t \to 0} \frac{n_{O_2}(t+\Delta t) - n_{O_2}(t)}{\Delta t} = 0.5 A_s \cdot J_{H_2}(t, n_{O_2}) \quad (6)$$

is susceptible to the original value $n_{O_2}(t=0) = 0.5 n_{O_2,0}$, where the $n_{O_2,0}$ is the $O_2$ amount inside HPM obtained from the previous one. This is a paragon of initial-value ordinary differential equation problem and is solved using a Runge–Kutta method of the fourth order with a variable step.

For $H_2$ evolution in the solar-concentrated heated HPM reactor, the first-law thermodynamic efficiency is defined as

$$\eta_{\text{HHV}} = \frac{n_{H_2} \cdot \text{HHV}_{H_2}}{\eta_{\text{abs}}^{-1}(Q_{H_2O} + Q_{\text{th}}) + \eta_{s \to e}^{-1}(n_{H_2O} W_{p1} \eta_{\text{fed}}^{-1} + n_{H_2} W_{p2} \eta_{\text{vacu}}^{-1} + n_{H_2O} W_{\text{sep}, H_2O})} \quad (7)$$

where $n_{H_2}$ and $n_{H_2O}$ are the molar amount of hydrogen produced and water input. $\text{HHV}_{H_2}$ is the molar higher heating value of the $H_2$. $\eta_{\text{abs}}$ is the absorption efficiency of the solar cavity receiver that the HPM reactor was set in and is a function of both $T_H$ and concentration level C of solar

energy [15]:

$$\eta_{abs} = 1 - \frac{\sigma \cdot T_H^4}{I \cdot C} \quad (8)$$

where σ is Stefan-Boltzmann constant, $I$ is the solar irradiation at Earth's surface and is conventionally taken as 1000 W/m². The solar concentration level $C$, for this study, is assumed to be $C$=5000 [15], and $\eta_{s \to e}$ is the conversion efficiency from solar radiation energy to electricity, assumed to be 0.15 and 0.4 throughout this study for characterization of different operating conditions (the former one is the commercial solar-to-electricity conversion efficiency of monocrystalline silicon, and the other is the solar-to-electricity conversion efficiency of multi-cell gallium arsenide cells in the laboratory.), respectively; $\eta_{fed}$ and $\eta_{vacu}$ are the efficiency of pump work (the ratio of exergy output to electricity input) for maintaining the feeding gas pressure and permeation H₂ pressure (single-stage vacuum pump), respectively, and can be defined as

$$\eta_{fed} = 85\%, \quad \eta_{vacu} = P_{H_2,perm}^{0.544} \quad (9)$$

and $Q_{H_2O}$ is the total heat input for raising the temperature of an amount of $n_{H_2O}$ moles of water from 25 °C to $T_H$, defined as

$$Q_{H_2O} = n_{H_2O} \left[ \int_{25°C}^{100°C} C_{p,H_2O(l)} dT + 40.872 + \int_{100°C}^{T_H} C_{P,H_2O(g)} dT \right] \quad (10)$$

where 40.872 kJ/mol is the molar phase change latent heat of water for transforming from liquid to gas phase and $Q_{th}$ is the total heat, which is need to be absorbed by water for its splitting, expressed as

$$Q_{th} = \Delta H \cdot \Sigma_{i=1}^{N} n_{i,H_2O} \quad (11)$$

where $\Delta H$ is the molar enthalpy change of the splitting reaction of the water, N= L/dL is the number of control volume, and $n_{i,H_2O}$ is the molar amount of reactant gas dissociated in the control volume $i$; $W_{p1}$ and $W_{p2}$ are the pump work consumed to feed the reactant gas into the high-pressure HPM tube, and that consumed to remove H₂ from the low-pressure chamber outside the HPM, respectively, and can be expressed as

$$W_{p1} = RT \ln(P_{H_2O}/P^{\ominus}), \quad W_{p2} = RT \ln(P_{H_2,ret}/P_{H_2,perm}) \quad (12)$$

where $P_{H_2O}$ is the pressure of the reactant gas within the OPM, $P^{\ominus}$ is standard pressure, R is universal gas constant, and $T$ is temperature in Kelvin; $W_{sep,H_2O}$ is the work consumed to separate 1 mol of unreacted water vapor from the mixed gas (including fuel and reactant gas) in the HPM tube at the outlet, which is negligible for the condensation of steam at room temperature (25 °C). The first-law thermodynamic efficiency as defined in Eq. 7 encompass the energy costs of heating the water (Eq. 10), endothermic enthalpy change of water-splitting (Eq. 11), pump work loss of feeding reactant gas in HPM tube (Eq. 12 left), vacuum pump work for providing low pressure to separate hydrogen outside (permeation side) HPM tube (Eq. 12 right) and re-radiation energy loss from the solar thermal collector that the HPM reactor sits in (Eq. 8).

Furthermore, to characterize the performance of HPM reactor, we defined the net solar-to-chemical efficiency

$$\eta_{net} = \frac{n_{H_2} \text{HHV}_{H_2} - W_p / \eta_{chem\text{-}mech}}{Q_{H_2O} + Q_{th}} \quad (13)$$

where $\eta_{chem\text{-}mech}$ is the efficiency for mechanical work being converted from chemical energy and considered as 40% [8], other variables are the same as before. All thermodynamic properties are calculated with HSC software [30].

## 3. Results and discussion

### 3.1 Analysis for partial pressures of gas products

To characterize and analyze the performance of HPM reactor, the numerical simulation was conducted in the theoretical model mentioned above. To begin with, we studied the basic state of water splitting, where the feeding gas rates is 1000 cm³/s and temperature is $T_H$=1500 °C. Under current conditions, we set the internal diameter and length of the tube to be 0.4 cm and 0.05cm (relatively short for the repaid $H_2$ permeation rate of Pd membrane) without the consideration of the warming-up length, respectively, which is enough for the reaction to reach steady state. However, the length of the tube varies with the flow rate, and the internal reaction of the tube is related to the percentage of the length. Therefore, the length is non-dimensionalized and taken as the percentage of length, for broad applicability at different operation conditions. The hydrogen partial pressure at permeation side is taken as $P_{H_2,perm} = 10^{-5}$ atm unless stated otherwise. We will discuss the gas ($H_2$, $O_2$ and $H_2O$) partial pressure and $H_2$ flux changing as they go through the tube at the gas feeding direction, in order to characterize the dynamics of the system. In addition, the thermodynamic efficiencies (first-law thermodynamic efficiency and net solar-to-chemical efficiency) and water conversion rate will be analyzed under different operating condition, such as different photoelectric conversion efficiency (0.15, 0.4) and disparate feeding gas pressure.

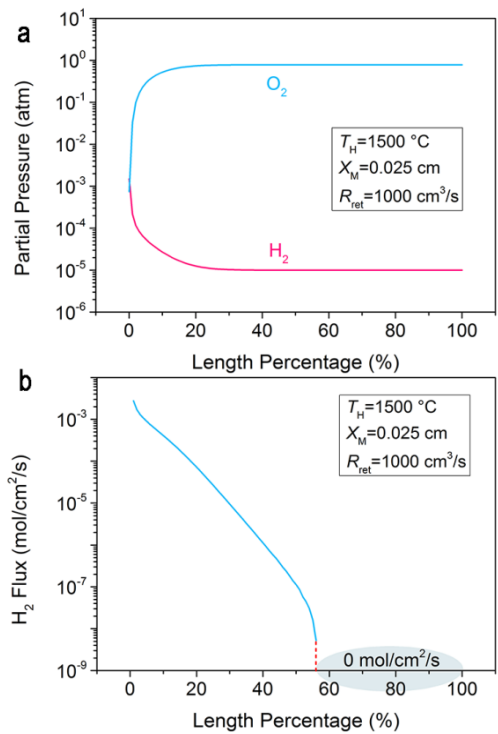

**Figure 3.** The characterization of basic reaction state, where reaction temperature is 1500 ℃, water flux $R_{ret}$ is 1000 cm³/s, internal diameter is 0.4 cm and membrane thickness is 0.025cm. (a) The partial pressure of hydrogen and oxygen in different positions of the reaction tube. (b) The $H_2$ flux in different positions of the reaction tube, gray area stands for the interval without $H_2$ flux.

The hydrogen permeation enhanced by vacuum pump is demonstrated in Fig. 3 and it shows that the hydrogen partial pressure and oxygen partial pressure started as $1.49 \times 10^{-3}$ atm and $7.45 \times 10^{-4}$ atm, which ratio is about 2:1, at the beginning of the journey of feeding gas in HPM reactor, due to the thermolysis of water at $T_H$. Then, as the reactant gas pushed unremittingly along the tube, $H_2$ is simultaneously extracted from the control volume through the HPM and thus the hydrogen partial pressure reduces, which would induce the equilibrium of water splitting shifts forward. Therefore, as the hydrogen is continuously separated (to the side of $10^{-5}$ atm), the water will continuously decompose and produce hydrogen and oxygen through the pipe. Because of this process, we can see in Fig. 3a the hydrogen partial pressure decrease and oxygen partial pressure increase monotonically, and eventually approach to the limit ($10^{-5}$ atm for hydrogen and 0.78 atm for oxygen). Besides, the driving force of hydrogen permeation reduced due to the decreasing difference of partial pressure of hydrogen between the two sides of HPM. Corresponding to this, the hydrogen permeation flux in Fig. 3b shows a virtually logarithmic linear decrease, because of the gradual decreasing hydrogen partial pressure difference between two sides of HPM, due to the permeation. After 54% of the tube length (gray area in Fig. 3b), the $H_2$ partial pressure on the reactant side reaches to $10^{-5}$ atm and thus the differential pressure inside and outside the membrane is balanced, so there is no more hydrogen through the HPM and the $H_2$ flux is zero.

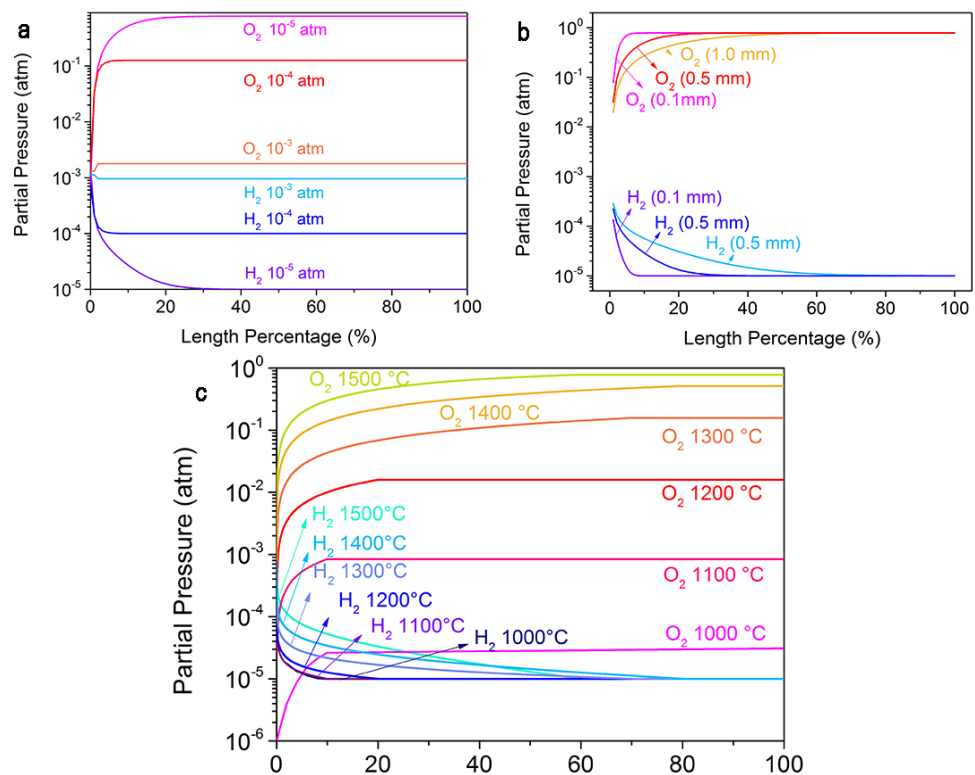

**Figure 4.** The variation of reactant partial pressures ($H_2$ and $O_2$) with the influence of negative pressure (permeation side), membrane thickness and reaction temperature. (a) The partial pressure of $H_2$ and $O_2$ in different positions of the reaction tube under different pressure out of HPM, at 1500℃, thickness of 0.025 cm and water flux of 1000 cm$^3$/s. (b) The partial pressure of $H_2$ and $O_2$ in different positions of the reaction tube with different thickness of hydrogen permeable membrane, at 1500℃, $10^{-5}$ atm permeation pressure and water flux of

1000 cm$^3$/s. (c) The partial pressure of H$_2$ and O$_2$ in different positions of the reaction tube under different reaction temperature, with thickness of 0.025 cm, 10$^{-5}$ atm permeation pressure and water flux of 1000 cm$^3$/s.

To further characterize the factors which can influence the performance of HPM reactor, we take the negative pressure (permeation side), membrane thickness and reaction temperature $T_H$ into consideration. The tube length is still 0.05 cm. For different negative pressure, in Fig. 4a, the hydrogen and oxygen partial pressure started at $1.49 \times 10^{-3}$ atm and $7.45 \times 10^{-4}$ atm, at 1500 $^o$C and 10$^{-5}$ atm of permeation side, and then increase and decrease monotonically to the limit (10$^{-5}$ atm for hydrogen and 0.78 atm for oxygen), which is same as the reaction in Fig. 4a. As we elevate the absolute value of negative pressure, the rate of increase of oxygen partial pressure and the rate of decrease of hydrogen partial pressure both reduce (Fig. 4a). This is because that the difference of hydrogen pressure between reactant side and permeation side decreases due to the increasing of permeation pressure, which thus compromises the driving force of hydrogen permeation, according to Eq. 3. However, the decreasing difference of hydrogen pressure between HPM indicates that the partial pressure will reach the limit pressure with extraction of less hydrogen. Therefore, the higher the permeation pressure provided by the vacuum pump is, the sooner (at smaller length of the tube) the partial pressure of hydrogen and oxygen reaches the limit. Despite the same trend, the partial pressure of hydrogen and oxygen reached higher and lower limit, respectively, as the negative pressure increasing. Specifically, for 10$^{-4}$ atm, the limits are 10$^{-4}$ atm and 0.126 atm for hydrogen and oxygen; for 10$^{-3}$ atm, the limits are 10$^{-3}$ atm and $1.79 \times 10^{-3}$ atm for hydrogen and oxygen, respectively.

According to Eq.3, the hydrogen permeation flux is negative proportional to the thickness of HPM ($X_M$). Therefore, as the membrane thickness increases, with no variation of other variables, the hydrogen transport flux will decrease. As a result, the rate of increase in partial pressure of oxygen and the rate of decrease in the partial pressure of hydrogen will both decrease, which can also be demonstrated in the absolute value of the slope of the curves in Fig. 4b. The partial pressures of hydrogen and oxygen reach their respective limits (10$^{-5}$ atm and 0.78 atm for H$_2$, O$_2$) more slowly, with increasing thickness. This fully shows the thickness is an important kinetic factor which controls the hydrogen permeation flux and hydrogen partial pressure at every moment.

The effect of reaction temperature is also analyzed in Fig. 4c, and it is a more complicated factor.

To explicate it, both the hydrogen permeable rate (Eq. 3) and the equilibrium constant (Eq. 1) are positively correlated with temperature. The conversion rate is positively correlated with the equilibrium constant, due to the Le Chatelier Principle. As seen from the Fig. 4c, the time for the reaction to reach equilibrium decreases as the conversion rate increases. From 1000 °C to 1400 °C, the length percentage where the gas partial pressure reaches limit increases with the temperature monotonically. As the temperature rises, the water conversion rate increases, and more hydrogen needs to be separated to reach the ultimate pressure of $10^{-5}$ atm. Therefore, the reaction gas needs to travel a longer distance in the HPM tube to reach the limit. In this process, conversion rate increases faster with temperature than hydrogen permeability rate does and it hence dominates the monotonically increasing critical length percentage. However, when temperature reaches 1500 °C, the critical length percentage decreases to 56%, which is lower than the length percentage of 1300 °C and 1400 °C. This manifests that hydrogen and oxygen partial pressures reach the partial pressure limits faster. Just like the analysis above, under this temperature, in spite of the increased conversion rate, the amplitude of increase in hydrogen penetration rate was greater and exceeded the amplitude of increase in conversion rate. As a result, the high hydrogen permeability rate dominates this process, so hydrogen and oxygen reach the partial pressure limit earlier.

*3.2 Analysis for water conversion rate*

Despite the different kinetic performance with different kinetic factor analyzed before, it is still the thermodynamic limit that really determines the performance of the system, which is also the final state of the kinetic reaction. Thereby, we make zero-dimensional thermodynamic calculation below and analyze the HPM system as a whole. We also conducted a similar calculation of conversion for OPM system (Fig. 5b) for comparison. In the same situation, the conversion rate of OPM is much lower than that of HPM. Specifically, in Fig. 5a, the water conversion rate increase monotonically with the temperature, due to the positive correlation between equilibrium constant and temperature [15]. At the same time, the conversion rate is also negative related to permeation pressure. This is collaborated by the Eq. 1 that the lowering pressure of hydrogen permeation allows the chemical equilibrium shift forward further to increase the conversion rate. Thanks to that hydrogen occupies twice the stoichiometric amount of oxygen, we got a much higher conversion rate than OPM. The highest conversion rate of 98% appears when temperature is 1800 °C and permeation pressure is $10^{-5}$ atm. Even for 1500 °C and $10^{-5}$ atm, the conversion rate is as appreciable as 87.8%, which is much higher than 1.26% for OPM (Fig. 5b).

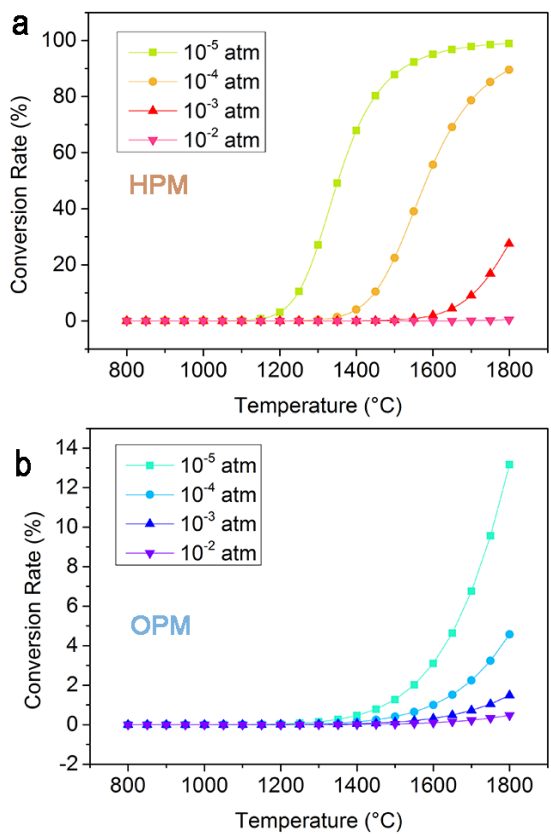

**Figure 5.** Water conversion rate versus temperature at different permeation pressures. (a) HPM reactor. (b) OPM reactor.

We also take the hydrogen separation ratio into consideration and delve its influence on conversion rate and hydrogen partial pressure (permeation side), and the calculation results are shown in Fig. 6, which is in thermodynamic equilibrium condition. It is obvious that, as the ratio of separated hydrogen increases, the equilibrium of water splitting reaction shifts forward, and the water conversion rate monotonously increases with the hydrogen partial pressure decreasing (the hydrogen partial pressures between the HPM are equal at equilibrium condition, irrespective of the pressure drop). Besides, the conversion rate and hydrogen partial pressure also increase with temperature, consistent with the principle of thermodynamic and chemical equilibrium above.

Therefore, increasing the proportion of hydrogen separation can promote a positive equilibrium shift and increase water conversion, but it will require a lower hydrogen partial pressure at the same time.

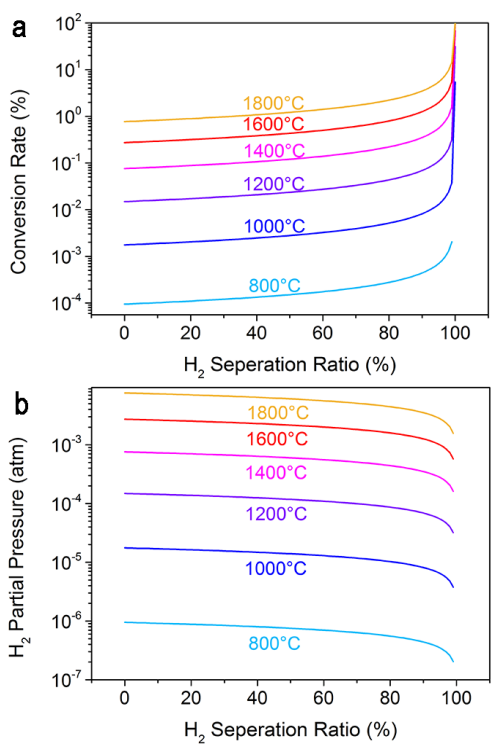

**Figure 6.** Conversion rate (a) and hydrogen partial pressure (b) versus hydrogen separation rate.

To further understand the thermodynamic relationship of water splitting, the relation between the reaction temperature and permeation pressure required for water splitting to reach a certain conversion rate (e.g. 90%) is also analyzed. In Fig. 7a, the positive correlation between the partial pressure of hydrogen and the reaction temperature is obvious, and the partial pressure of hydrogen also increases with the decrease in the conversion rate that needs to be reached. For example, to reach a 70% conversion rate (the blue line), the partial pressure is needed to reach $4.41 \times 10^{-9}$ atm at 900 °C, but, when temperature elevates to 1500 °C, the partial pressure is only to approach $3.01 \times 10^{-4}$ atm, which is more practical and efficient for vacuum pump to maintain this pressure. Therefore, in actual production, we need to consider comprehensively the temperature and the partial pressure of hydrogen to achieve a conversion rate that we want. In comparison, for OPM reactor, we can clearly see that reaching the same conversion rate requires lower oxygen partial pressure than HPM. Lower absolute value of partial permeation pressure will reduce vacuum pump efficiency, resulting in more energy consumption. This reflects the advantages of HPM's high conversion rate (Fig. 7b).

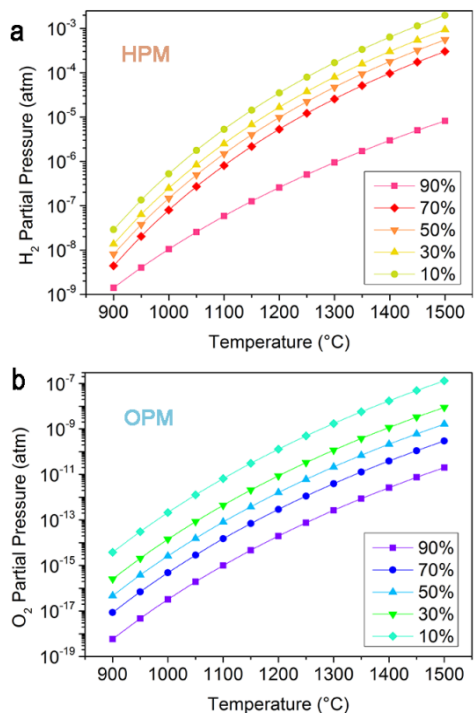

**Figure 7.** The relationship between the critical partial pressure and temperature when conversion rate is fixed. (a) HPM reactor. (b) OPM reactor.

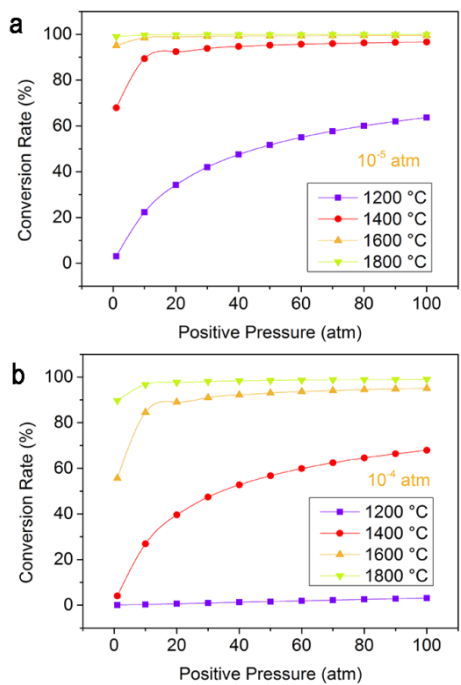

**Figure 8.** The characterization of conversion rate when exert positive pressure (reactant pressure). (a, b) The relation between conversion rate and positive pressure for different temperature under the negative pressure $10^{-5}$, $10^{-4}$ atm, respectively.

To step further, we also analyze the operation condition where positive pressure (feeding gas pressure) is taken into consideration, which is what the previous OPM researches did not do [5, 15]. According to Eq. 3, increasing feeding gas pressure can increase the conversion rate. This is because that the equilibrium would be shifted forward by exerting higher feeding gas pressure, and therefore more $H_2$ would be separated outside the HPM reactor. As we can see, the conversion rate increases simultaneously with the reaction temperature and feeding gas pressure. The highest conversion rate (99.90%) appears at 1800 °C, $10^{-5}$ atm permeation pressure and 100 atm react pressure (Fig. 8a, b). Our calculation range is only in the range of 1-100 atm, because the pressure in the high temperature chemical process is generally much less than 10 MPa.

*3.3 Thermodynamic efficiency analysis*

Except the conversion rate, the efficiency is another important factor to characterize the performance of the HPM water-splitting reactor. Therefore, we define two different efficiencies: first-law thermodynamic efficiency (Eq. 7) and net solar-to-chemical efficiency (Eq. 13). These efficiencies are calculated in different operation conditions and their analyses would be conducted below.

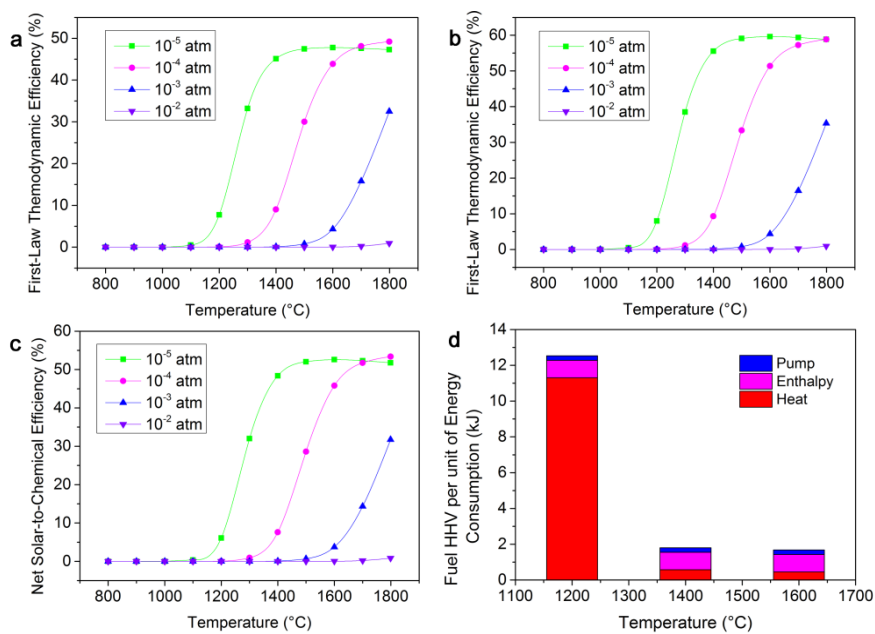

**Figure 9.** Efficiency analysis without exerting feeding gas pressure. (a, b) The relation between efficiencies of the system and reaction temperatures under different hydrogen permeation pressures with the 0.15 and 0.4 solar-to-electric efficiency, respectively. (c) Net solar-to-electric efficiency for different permeation pressure at different temperature. (d) Energy costs for the production of 1 kJ of hydrogen by HPM reactor at different temperature for $10^{-5}$ atm permeation pressure.

---

To begin with, the most usual condition (without feeding gas pressure) is analyzed. Here, we consider the different solar-to-electric efficiency (0.15 and 0.4), which is used to convert electricity (vacuum pump work) into solar energy, and then with ideal separation exergy which is calculated in most of literatures published [5, 15]. In order, for the high permeable hydrogen pressure of $10^{-2}$ atm (or other higher pressures), the first-law thermodynamic efficiencies and net solar-to-chemical efficiencies are almost negligible (Fig. 9a, b, c). For $10^{-3}$ atm and $10^{-4}$ atm permeation pressures, they increase monotonically with temperature. In addition, they rise quickly first and then grow more slowly with increasing temperature for $10^{-4}$ and $10^{-5}$ atm and reach the top value about 58.5% and 59.1% with the solar-to-electric efficiency of 0.4 (Fig. 9b), respectively, and the net solar-to-chemical efficiency is as high as 53.4% at 1800 °C and $10^{-4}$ atm permeation pressure. This trend has been explained in Fig. 9d, which is the illustration of energy costs for the production of 1 kJ of hydrogen by HPM reactor (the pump work in Fig. 9d is converted into solar energy with solar-to-electric efficiency of 0.4). The reciprocal of the vertical axis number in Fig. 9d is the first-law thermodynamic efficiency in Fig. 9b, and the trend of lines in Fig. 9a, c are similar with those of Fig. 9b, so they won't be covered here. Under $10^{-5}$ atm permeation pressure, the energy cost to heat water to produce 1kJ $H_2$ (red area) decrease significantly from 1200 °C to 1400 °C. However, from 1400 °C to 1600 °C, the heating water energy is almost the same. Besides, other factors change a little with temperature. To explicate this, the enthalpy is the same for producing 1kJ $H_2$ ($\frac{q_{th}}{HHV}$, $q_{th}$ is the enthalpy for 1 mol water-splitting reaction), but the energy cost to heat water and pump loss varies with positive pressure. To elucidate the relation between the conversion rate and energy cost to heat water (heat energy), we need to analyze from the heat energy formula,

$$Q_{H_2O,per} = \frac{n_{H_2O}}{n_{H_2}} \cdot \frac{q_{H_2O}}{HHV} = \frac{1}{\alpha} \cdot \frac{q_{H_2O}}{HHV} \tag{14}$$

where the $q_{H_2O}$ and $\alpha$ are the heat to rise the water temperature (varies little with the temperature) and conversion rate, respectively. This clear demonstrates that the higher the conversion rate is, the lower heat energy cost to heat water for producing 1 kJ $H_2$. Thus, the significant decrease of heat energy from 1200 °C to 1400 °C can be explained by the significant increase from 1200 °C to 1400 °C of conversion rate (Fig. 5a). The little change from 1400 °C to 1600 °C is also explained in the same way. With the fixed pump operation temperature (25 °C), enthalpy for producing 1kJ $H_2$ ($\frac{q_{th}}{HHV}$), and little-changed $q_{H_2O}$, the trend of the first-law thermodynamic efficiency under $10^{-5}$ atm is explained by clearly demonstrating different energy cost factors. Such trend shows the importance of conversion rate for first-law thermodynamic efficiency. By doing so, the trends of other permeation pressure efficiencies can also be explicated in a similar method.

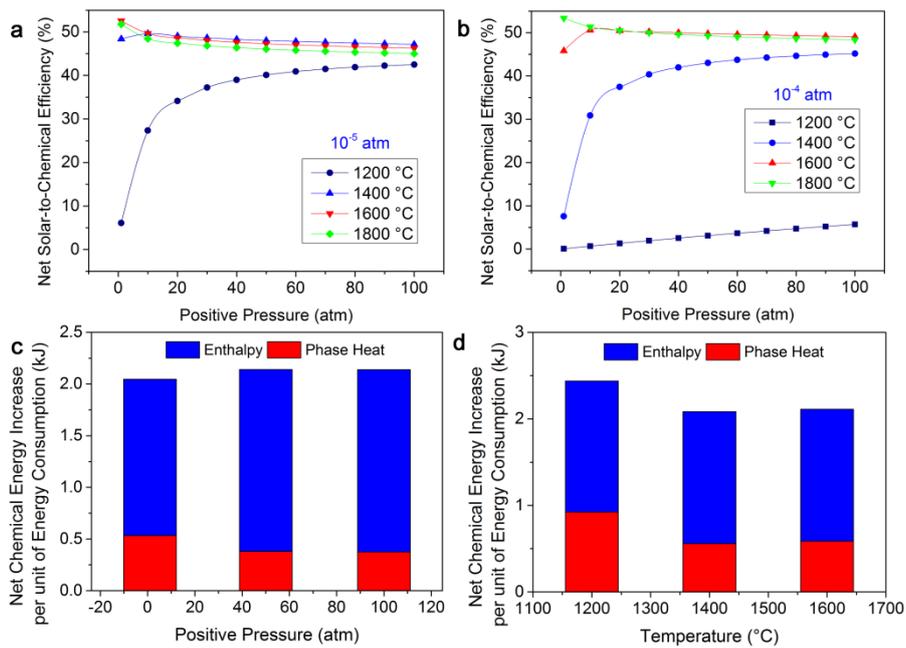

**Figure 10.** The characterization of efficiency when exert positive pressure (reactant pressure). (a, b) Efficiency versus positive pressure for different temperature under the negative pressure $10^{-5}$, $10^{-4}$ atm, respectively. (c) Energy costs for the production of 1 kJ of hydrogen by HPM reactor at different positive pressure for $10^{-5}$ atm permeation pressure at 1400 °C. (d) Energy costs for the production of 1 kJ of hydrogen by HPM reactor at different temperature for $10^{-5}$ atm permeation pressure and 50 atm positive pressure.

Encouraged by the improved conversion rate with the help of positive pressure (Fig. 8), the net solar-to-chemical efficiency is calculated when the positive pressure is exerted (Fig. 10a, b). Different from the monotonically increasing trend of conversion rate with positive pressure, the trend of net solar-to-chemical efficiency is a little complicated, so the energy analysis histograms (Fig. 10c, d) are also necessary to demonstrate every energy cost factor in the system. We can see that, for $10^{-5}$ atm permeation pressure, the solar-to-chemical efficiency varies a little with the increasing positive pressure under 1400 $^{o}$C, 1600 $^{o}$C and 1800 $^{o}$C. Take the 1400 $^{o}$C for an example (Fig. 10c), the enthalpy and phase heat (heating water energy cost) for 1 atm feeding gas pressure are both lower a little than those for 50 atm and 100 atm feeding gas pressure, due to the lower pump work loss. Thus, the solar-to-chemical efficiency is a bit higher under 1 atm than 50 atm 100 atm feeding gas pressure. Besides, these two energy cost factors are almost the same for 50 atm and 100 atm, under 1400 $^{o}$C and therefore the net solar-to-chemical efficiency is almost the same. The trends of other temperatures can also be explained in this way.

In addition, the relation between net solar-to-chemical efficiency and temperature, when 50 atm positive pressure and $10^{-5}$ atm permeation pressure are exerted, is analyzed by Fig. 10d. As we can see, the enthalpy cost is almost them same for different temperature but the phase heat decreases significantly from 1200 $^{o}$C to 1400 $^{o}$C. The decreasing phase heat is correlated with the increasing conversion rate (Fig. 8a), which can be explained by Eq. 14. Therefore, the net solar-to-chemical efficiency enhances a lot from 1200 $^{o}$C to 1400 $^{o}$C. The explanation for the little-changed net solar-to-chemical efficiency is similar to this. The highest efficiency of 53.4% appears at 1 atm feeding gas pressure, $10^{-4}$ atm permeation pressure and 1800 $^{o}$C. With the help of energy cost factors diagrams, the trends for net solar-to-chemical efficiency are clearly demonstrated.

However, these two efficiencies (first-law thermodynamic and net solar-to-chemical efficiency) are very ideal. In practical production, mechanical vacuum pump efficiency decreases sharply with decreasing permeation pressure (Eq. 9). Thus, we also demonstrated the first-law thermodynamic efficiency with real pump work efficiency (the ratio of separation exergy to electricity consumed) (Eq. 9), which has guiding significance for the choice of vacuum pump in actual production. The highest first-law thermodynamic efficiencies are 5.05% and 11% with the vacuum pump efficiency and solar-to-electric efficiency of 0.15 and 0.4, respectively, at permeation pressure of $10^{-3}$ atm and 1800 $^{o}$C. For 1500 $^{o}$C, a temperature which is more commonly used in focusing high temperature solar energy [32, 33], the highest efficiencies are 1.21% and 3.05% at $10^{-4}$ atm for solar-to-electric efficiency of 0.15 and 0.4, respectively. This exceeds far the experimental efficiency of the dual-temperature method, which is only a few thousandths [6]. It can be clearly seen that although the lower hydrogen permeability pressure will increase the conversion rate, but it will also significantly reduce the vacuum pump efficiency, resulting in a decline in first-law thermodynamic efficiency. In actual production, the key factor that restricts the efficiencies is whether or not an efficient way to separate gases can be found.

## 4. Conclusion

In conclusion, HPM was introduced into a thermochemical water-splitting system and established a theoretical numerical model to characterize various aspects of its performance for the first time. The practical reaction along the HPM tube's one-dimensional working condition in different environments (different HPM thicknesses, different permeation pressures and different reaction temperatures) was simulated to reflect the actual production conditions as realistically as possible. Encouraged by the chemical equilibrium theory, a conversion rate of 87.8% was achieved at $10^{-5}$ atm permeation pressure and 1500 $^{o}$C, which is much higher than the 1.26% of the OPM system under the same conditions. Furthermore, the appreciable first-law thermodynamic efficiency of 59.1% and net solar-to-chemical efficiency of 53.37% have been realized. When the mechanical pump efficiency is considered, the more practical first-law thermodynamic efficiency equals 3.05%, which overweight the OPM efficiency of a few thousandths profoundly. Besides, the effect of feeding gas pressure on the system was also analyzed. As predicted by thermodynamics, adding feeding gas pressure will cause the equilibrium to shift forward and increase the conversion rate (even reach a high value of 99.90%). However, an excessively high positive pressure will consume more pump work and eventually lead to a drop in system efficiency. This study provides detailed theoretical guidance for the actual production of this new and efficient HPM water splitting reactor, and paves the way of solar water-splitting by membrane reactor, which might play a significant role in the field of solar-to-hydrogen conversion on a large scale.

**Acknowledgement**

We thank Liming Zhao (a third year undergraduate from Department of Materials Science and Engineering, Southern University of Science and Technology) for the assistance in the schematic drawing, Xixi Hu (a third year undergraduate from School of Control Science and Engineering, Shandong University) and Yicheng Xue (a third year undergraduate from School of Electrical Engineering, Shandong University) for helping of calculation, and Jinwei Xu (Ph.D from Department of Materials Science and Engineering, Stanford University) for data art design.

**Highlights**

1. Solar thermochemical water-splitting reaction with hydrogen permeation membrane reactor is first analyzed.
2. High conversion rate of 87.8% for usual operation condition is realized.
3. Appreciable first-law thermodynamic efficiency of 59.1% is achieved.
4. Feeding gas pressure was considered and the net solar-to-chemical efficiency of 53.37% is realized.

**Nomenclature**

| | |
|---|---|
| $A_s$ | Side area of the control volume |
| C | Concentration level of solar energy |
| $C_{p,H_2O(l)}$ | Heat capacity for liquid water |
| $C_{p,H_2O(g)}$ | Heat capacity for vapor |
| $dL$ | Length of every single control volume |
| HHV | Higher heating value of $H_2$ |
| $I$ | Solar irradiation at Earth's surface |
| $J_{H_2}$ | Flux of $H_2$ |
| K(*T*) | Chemical equilibrium constant |
| *k* | Sieverts constant |
| L | Total length of HPM tube |

| Symbol | Description |
|---|---|
| $N$ | The number of control volume |
| $n_{H_2}$ | Amount of $H_2$ |
| $n_{H_2O}$ | Amount of water input |
| $n_{i,H_2O}$ | Molar amount of reactant gas dissociated in control volume $i$ |
| $n_{O_2}$ | Amount of $O_2$ |
| $P_{H_2}$ | Partial pressure of $H_2$ |
| $P_{O_2}$ | Partial pressure of $O_2$ |
| $P_{H_2O}$ | Partial pressure of $H_2O$ |
| $P_{H_2,\text{ret}}$ | $H_2$ partial pressure inside (reaction side) the HPM tube |
| $P_{H_2,\text{perm}}$ | $H_2$ partial pressure outside (permeation side) the HPM tube |
| $P^{\ominus}$ | Standard pressure |
| $Q_{H_2O}$ | Total heat input for raising the temperature to $T_H$ |
| $Q_{\text{solar}}$ | Solar energy input to produce $H_2$ |
| $Q_{\text{th}}$ | Total heat need to be absorbed by water for its splitting isothermally |
| $q_{\text{th}}$ | Heat for 1 mol water-splitting reaction |
| $q_{H_2O}$ | Heat to rise 1 mol water temperature to $T_H$ |
| $T_H$ | Temperature for water-splitting reaction |
| $T_L$ | Room temperature |
| $W_{p1}$ | Pump work for pump work for maintaining the feeding gas pressure |
| $W_{p2}$ | Pump work for pump work for maintaining the permeation gas pressure |
| $X_M$ | Thickness of HPM |
| $\Delta H$ | Molar enthalpy change of the splitting reaction of the water |
| $\alpha$ | Water conversion rate |
| $\eta$ | Thermodynamic efficiency |
| $\eta_{\text{abs}}$ | Absorption efficiency of the solar cavity receiver |
| $\eta_{\text{chem-mech}}$ | Efficiency for mechanical work being converted from chemical energy |
| $\eta_{\text{fed}}$ | Efficiency of pump work for maintaining the feeding gas pressure |
| $\eta_{\text{HHV}}$ | First-law thermodynamic efficiency |
| $\eta_{\text{net}}$ | Net solar-to-electric efficiency |
| $\eta_{\text{pump}}$ | Efficiency of vacuum pump |
| $\eta_{s \rightarrow e}$ | Solar-to-electric conversion efficiency |
| $\eta_{\text{vacu}}$ | Efficiency of pump work for maintaining the permeation gas pressure |
| $\sigma$ | Stefan-Boltzmann constant |